# The relationships among GC content, nucleosome occupancy, and exon size

**Liya Wang[1][§], Lincoln D Stein[2, 3], Doreen Ware[1, 4]**

[1]Cold Spring Harbor Laboratory, Williams 5, Cold Spring Harbor, NY 11724

[2]Ontario Institute for Cancer Research, 101 College St. Suite 800, Toronto, ON, M5G0A3, Canada

[3]Department of Molecular Genetics, University of Toronto, 1 King's College Circle, Toronto, ON MSS 1AB, Canada

[4]US Department of Agriculture/Agricultural Research Service North Atlantic Area, Robert W. Holley Center for Agriculture and Health, Tower Road, Ithaca, NY 14853

[§]Corresponding author

Email addresses:

    Wangli@cshl.edu

    Lincoln.Stein@gmail.com

    Ware@cshl.edu

# Abstract

**Background**

The average size of internal translated exons, ranging from 120 to 165 nt across metazoans, is approximately the size of the typical mono-nucleosome (146 nt). Genome-wide study has also shown that nucleosome occupancy is significantly higher in exons than in introns, which might indicate that the evolution of the exon-intron structure is related to the chromatin organization.

**Results**

By grouping exons by the GC contents of their flanking introns, we show that the average exon size is positively correlated with its GC content. Using the sequencing data from direct mapping of *Homo sapiens* nucleosomes with limited nuclease digestion, we show that the level of nucleosome occupancy is also positively correlated with the exon GC content in a similar fashion. We further demonstrate that exon size is positively correlated with their nucleosome occupancy when GC content in their flanking introns is low.

**Conclusions**

The strong correlation between exon size and nucleosome occupancy suggests that chromatin structure plays an important role in the evolution of exon-intron structure.

# Background

The nucleosome is the fundamental building unit of eukaryotic chromatin, consisting of about 146 nucleotide (nt) pairs of double-stranded DNA wrapped around eight histone molecules [1]. It has become clear that the biological roles of nucleosomes extend far beyond simple DNA packaging to include transcriptional control, co-transcriptional splicing, DNA replication, recombination, and repair [2-5]. When DNA is wrapped around a histone octamer, it inhibits the binding of polymerase, regulatory proteins, DNA repair and recombination complexes, possibly due to steric hindrance. Nucleosomes are depleted in active regulatory sequences [6-8], but dense in coding sequence regions. In particular, nucleosome occupancy is markedly higher in exons versus introns [4, 9-12], possibly indicating that nucleosome provides a protective role in protein coding regions. It is also possible that nucleosome aids in splicing recognition and may play a significant role in alternative splicing. High nucleosome occupancy is thought to slow RNA polymerase II (Pol II) elongation [13, 14] so that Pol II can pause, unwind the DNA double strand to release it from the nucleosome. The transcriptional rate of Pol II may also play a role in cotranslational recognition of splicing signals in the pre-mRNA [15-17].

From an evolutionary perspective, the existence of introns in eukaryotic genes have been a mystery ever since their discovery [18-20]. The advantages of having introns may include generating new genes through exon shuffling [21], and producing multiple isoforms and translated proteins through alternative splicing [22]. However, it is not clear how the exon-intron structure evolved and what driving forces have shaped the exon-intron structure of current-day genomes over evolutionary time. In previous work [23], we have demonstrated that the size distribution of internal translated exons (or itexons, as shown in Figure 1) could be well fitted with a lognormal function. We also introduced a stochastic fragmentation process (GRFP) that can numerically reproduce such size distribution via simulation given total coding sequence length. However, the simulation is only good at estimating genome wide averages (such as intron density) by ignoring the internal property of DNA sequences, e.g., the GC content. In this study, we set out to explore the potential correlations among GC content, nucleosome occupancy, and exon size to extend our understanding of the evolution of exon-intron structures.

A wide variety of DNA sequence features are believed to affect nucleosome formation, including base composition (GC content) [9, 24, 25], nucleosome-excluding sequence especially poly-dA/dT tracts [24-29], and physical properties of DNA [30]. It has been shown that GC content positively dominates intrinsic nucleosome occupancy [31], with the next most notable factor being the frequency of AAAA. The dominating role of GC content is probably due to the facts that high GC content both reduces frequency of poly-A-like stretches and correlates with many other DNA structural characteristics. It has also been shown that nucleosome occupancy is contingent on the GC content difference between exons and their flanking introns [9, 32]. Intuitively, a nucleosome will be better positioned in regions of higher GC region if the flanking regions have lower GC content. Therefore, we examined the relationship between exon size and nucleosome occupancy in the context of GC content in exons and their flanking introns. Our analysis shows that there is a strong correlation between nucleosome occupancy and the GC architecture of the exon-intron structure. For groups of exons flanked by introns with similar GC content, our results highlight the strong correlations between exon size and nucleosome occupancy. The correlation might suggest a novel role of nucleosome in the evolution of exon-intron structure of eukaryotic genomes.

## Methods

### Data sets

The direct sequencing data of nucleosome ends using MNase digestion in *H. sapiens* activated CD4+ T cells [33] is downloaded from NIH website. All reads (25 nt) are mapped to the human genome (hg19) using short read aligner BWA [34]. A sliding window of 10 nt is applied across all chromosomes to generate the nucleosome profiles. For each window, all reads mapping to the sense strand 80 nt upstream of the window and reads mapping to the antisense strand 80 nt downstream of the window are counted.

For calculating nucleosome occupancy in introns, we average the scores in the first 60 nt in the upstream intron sequence and the last 60 nt in the downstream intron sequence (-120 nt to -70 nt in Figure 3). For nucleosome occupancy in exons, we average the scores between 30 nt and 60 nt. The nucleosome occupancy values shown in Figure 4 and 5 are calculated as the nucleosome occupancy difference between exon and intron. In Figure 3, for 3′ splice site, all reads mapping to the sense strand 80 nt upstream of the window are counted and then averaged across exons; for 5′ splice site, all reads mapping to the antisense strand 80 nt downstream of the window are counted and then averaged across exons. The resulted nucleosome occupancy profile is different from those generated by averaging both sense and antisense strand reads in previous studies [9, 33]. The major reason for calculating nucleosome occupancy this way is to build an occupancy profile that is sensitive to the exon-intron junctions. Since the sequence data captures both ends of the nucleosome binding fragment, such approach allows us to better capture one end of the fragment that covering the exon-intron boundaries. For example, for 3′ splice site (Figure 3), the nucleosome occupancy in intron will be overestimated if counting reads falling in the antisense strand downstream region, which are the upstream ends of the fragments covering the boundaries. Similarly, the nucleosome occupancy in exon will also be overestimated if counting these mapped sequence reads, which are the upstream ends of the fragments not covering the 3′ splice site. Another reason for our way of calculating nucleosome occupancy is to reduce effect of exon sizes - long exons tend to accumulate more nucleosomes than shorter ones. Such exon size effect can be minimized by counting reads falling in flanking intron regions, and exon regions only if they are close to splice junctions. As mentioned in the method section, we also excluded exons flanked by introns shorter than 200 nt, which greatly reduces the chance of counting nucleosome reads binding to adjacent exons.

The exon and flanking intron sequences of hg19 are downloaded from UCSC Table Browser [35] for calculating the exon size and GC content. To ensure the quality of the data, we used exons satisfying these criteria: exons in protein coding gene with both RefSeq mRNA ID and known status of both gene and transcript. Exons in genes with transmembrane or signal domains are also excluded because their size distribution does not follow a lognormal distribution (data not shown). Exons shorter than 25 nt are excluded to avoid GC content bias by the short size. 120 nt intron sequences on each side of the exon are used for comparison of GC content (and exons flanked by introns shorter than 200 nt are excluded). For the upstream intron sequence, we discarded three nt from the 3′ end as these are part of the splice-site signal. For the same reason, we also discarded the first six nt of the downstream introns and the first two and the last three nt of the exons for calculating GC content.

**Regression analysis**

We generate histograms for exons falling in each GC group and fit them with a lognormal distribution as done previously [23]:

$$dN = \left( N/(\sigma_E \sqrt{2\pi}) e^{-\left((\ln E - \mu_E)/\sqrt{2}\sigma_E\right)^2} \right) d\ln E$$

Where $E$ is exon size, $dN$ the number of exons (with a certain range of sizes) in a bin (bin size of the histogram is 0.1), $N$ the amplitudes of the peak; $\mu_E$ the mean position (or the average exon size for each group), and $\sigma_E$ the standard deviation of the lognormal distribution. These and subsequent fittings in this study are performed using the nonlinear Trust-Region-Reflective curve-fitting algorithm [36, 37].

## Results and Discussion

### Internal translated exons are constrained in size

Exons can be grouped into 12 mutually exclusive categories by containing what transcriptional or translational boundaries [38]. The size distributions of the eight most common classes are shown in Figure 1 and fitted with a lognormal function or a mixture of two lognormal functions: (1) an iuexon is an internal un-translated exon; (2) an itexon is an internal translated exon; (3) an iutexon is an internal exon having a 3′ portion of the 5′ UTR followed by a CDS; (4) an ituexon is an internal exon having a 5′ portion of 3′ UTR following a CDS; (5) a 5utexon is the 5′-terminal exon having a 5′-untranslated region (5′ UTR) followed by a coding sequence (CDS); (6) a 3tuexon is the 3′-terminal exon having a 3′ UTR following a CDS; (7) a 5uexon is the 5′-terminal untranslated exon in a gene; (8) a 3uexon is the 3′-terminal untranslated exon [38].

Figure 1 shows that internal exons (top four) are more constrained in sizes than others. Among them, itexons is the most constrained (in terms of standard deviation). The facts that itexons are also the most abundant ones (by counts) and can be well fitted with a lognormal function allow us to easily quantify how the mean size of them is correlated with both GC content and nucleosome occupancy in the following sections.

**Exon size increases with its GC content**

We first grouped exons by binning on increased GC content. For exons in each group, we fit the size distribution of the exons with a lognormal function (see methods section) and take the mean value of the lognormal function as the average exon size of the group.

For *H. sapiens*, its exon size increases with exon GC content (solid line in Figure 2), with the exception of exons with medium GC content (0.4 – 0.52), where the average exon size is nearly constant at around 120 nt (or 4.78 in log scale). To explore the contribution of flanking introns, we divide exons into five non-overlapping groups by GC content of their flanking introns (shown in the legend). In each group, exon size increases with exon GC content (dashed lines in Figure 2). Therefore, the constant size of exons with medium GC content is likely resulting from the averaging effect of flanking introns (with various GC contents). These observations indicate that the average exon size is a function of GC contents in exons and their flanking introns. Previous study [39] has also showed that exons with differential exon-intron GC content display higher nucleosome occupancy than flanking introns and other exons, which might indicate that the exon size is correlated with its nucleosome occupancy.

**Nucleosome occupancy marks exon-intron boundary**

Not like GC content or exon size that can be easily quantified, it is more ambiguous to quantify the level of nucleosome occupancy for an exon. For example, assuming binding site is within exon when possible, long exons (e.g., over 350 nt) could bind more than one mono-nucleosome [40], while exons with size of 250 nt could bind mono-nucleosome from zero nt to 100 nt relative to the 3′ splice site. Therefore, we made an assumption that the evolution of exon size, if related, is most likely to be affected by the mono-nucleosome that clearly marks exon-intron boundaries. Based on this assumption, we build the nucleosome occupancy profile for both splice sites of each itexon, as discussed in the methods section.

Figure 3 demonstrates that the genome-wide profile of nucleosome occupancy clearly marks exon-intron boundaries for both 3′ splice site (left) and 5′ splice site (right). Therefore, we can quantify how well the nucleosome occupancy marks the exon-intron boundary with the difference between the nucleosome occupancies in exon and that in intron (as shown in Figure 4 and Figure 5). Such quantification will allow us to examine its correlation with exon size or GC content in the following sections.

**Nucleosome occupancy increases with GC content**

Figure 2 demonstrates that the average exon size clearly increases with the exon GC content (if flanked by introns with similar GC content). Given that GC content plays a dominant role in intrinsic nucleosome occupancy [31], we next investigated whether nucleosome occupancy is related to the GC architecture of exon-intron structure in a similar fashion as the exon size.

To explore the relationship between nucleosome occupancy and exon GC content, we firstly grouped exons into five non-overlapping groups by the GC content of their flanking introns (shown in the legend of Figure 4). In each group, the correlation between the nucleosome occupancy and exon GC content is shown, with markers representing the average values and vertical bar representing the standard errors. Figure 4 shows that the nucleosome occupancy have a strong positive correlation with the exon GC content when flanked by introns with low GC content (<0.4). The correlation is weaker but still genrally positive for those flanked by introns with GC content from 0.4 to 0.6, then get stronger when flanked by introns with high GC content (from 0.6 to 0.7). It is worthwhile to point out that, by grouping exons by the dominating feature – GC content, other related DNA sequence features affecting intrinsic nucleosome occupancy are most likely averaged out. However, GC content is not directly related to the formation of nucleosome, which might partially explain why the level of nucleosome occupancy has a complicated relationship with exon GC content as shown in Figure 4.

Additionally, higher GC content in flanking introns might give introns higher chance to position nucleosome on themsevles, thus making exon-intron boundaries "blurred" for nucleosome positioning. Similar observations have been made on the decreased exon nucleosome occupancy when intron GC content is high [39]. However, when the GC content in flanking introns is very high ($> 0.6$), the level of nucleosome occupancy increases again, which is a novel phenomena and indicates that the exon-intron boundary marking property of nucleosome occupancy is restored.

**Exon size increases with nucleosome occupancy**
The above analyses demonstrate that both exon size and nucleosome occupancy are strongly related to the GC architecture of the exon-intron structure. Thus, it is highly likely that the exon size is related to the level of nucleosome occupancy. Figure 5 confirms the correlation between them, with the solid lines each representing exons flanked by introns with GC content ranging from 0.2 to 0.7 (bin size 0.1). For each group of exons, nucleosome occupancy values are used to further group them and a lognormal function is used to fit the exon size distribution. The average exon size and standard deviation are calculated as the mean and the standard deviation of the lognormal function.

Figure 5 shows that the exon size increases with the level of nucleosome occupancy on itself. As GC content in flanking introns increases, exon size becomes less sensitive to nucleosome occupancy, which manifests itself as a decrease of the slope of the curve. The decreased sensitivity is roughly consistent with the observation on how nucleosome occupancy is correlated with GC content (Figure 4). Furthermore, combining the results shown in Figure 2 and 5, it is clear that exon size is more related to nucleosome occupancy when the GC content in their flanking intron is low (<0.4), where nucleosome can be better positioned within exons. On the contrary, when the intron GC content is above 0.4, exon size is more related to its own GC content instead of nucleosome occupancy, which agrees with previous observation that nucleosome occupancy can not mark exon-intron boudary at high intron GC content [39].

Interestingly, Figure 5 shows that exons flanked introns with low GC content (from 0.2 to 0.3) tend to reach 146 nt (dashed line) in the average size given sufficient high nucleosome occupancy, which is roughly the size of the DNA fragment that can wrap around a mono-nucleosome. Based on these observations, we hypothesize that dense nucleosome occupancy on exon might play a role in constraining the exon sizes over evolution.

**Losing nucleosome marking results in shorter and less constrained exons**
By grouping internal translated exons by ordinary positions, we have observed that exons near TSS shows larger standard deviation than those in the middle of the gene and those near Transcription End Site (TES) [23]. This phenoma held for all vertebrate genomes we investigated. On the other side, phased nucleome bindings has been observed on both sides of TSS for expressed genes [33]. To confirm the correlation with exon size, we grouped exons by their distance to TSS and TES and calculated mean, standard deviation (Figure 6), and the nucleosome occupancy for each group (Figure 7).

Not surprisingly, Figure 6 shows that exons near TSS tend to be shorter and less constrained (the larger standard deviation). Figure 7 shows that, for exons near TSS (within 500 nt), the level of nucleosome occupancy is much higher in regions near TSS (ignoring the exon-intron boundary), which indicates that this region is occupied by nucleosomes that are aligned relatively to TSS, instead of the exon-intron architectures. Therefore, it is possible that exons near TSS are not well marked by nucleosome occupancy, which, in turn, makes them significantly shorter and less constrained than those exons inside the gene body. In addition, Figure 6 shows that exons near TES are also shorter, which agrees with Figure 7 where exons near TES are not well marked by nucleosome occupancy (but not as significant as those near TSS).

**The equilibrium of intron gain and loss**
We have argued that the genome-wide intron gain/loss has reached equilibrium in vertebrate genomes, and the evidence is that intron density is almost the same for all well annotated vertebrate genomes [23]. Another hypothesis on exon size restriction is the exon definition model [41], which suggests that vertebrate exons need to be longer than 50 nt to avoid steric hindrance that blocks the binding of RNA-protein complexes on both sides of the exons. Based on the observation of empirical exon size distribution, the model also suggested that exons shall not be longer than 300 nt [41], and later study took a more conserved value of 250 nt as maximum exon size [42].

Based on above analysis and precious studies [9, 42], it might be also reasonable to argue that exon sizes are optimized for positioning mono-nucleosome. One related question will be what the minimum exon size is for effectively positioning a mono-nucleosome. To answer this, we create the nucleosome occupancy density map (Figure 8) through grouping exons by size from 30 nt to 330 nt with a bin size of 10 nt. The nucleosome occupancy along exons and their flanking introns are averaged to create the density map, which shows that, when exon is shorter than 60 nt, it is hard to distinguish exon from intron. This suggests that exon needs to be roughly at least 70 nt long to effectively position a mono-nucleosome.

With the GRFP model we proposed before [23], we can numerically estimate the effect of extra intron gains on the percentage of exons with the "optimal" sizes for both fitting the exon definition model and effectively positioning a mono-nucleosome. Taking *H. sapiens* genome as an example, there are 104115 internal translated exons with total length of 13256978 nt (after removing exons longer than 400 nt), giving an average intron density of 7.85/kb. Using the GRFP model, we can exactly reproduce the empirical size distribution of the exons by inserting 104114 introns into one artificial exon of length 13256978 nt. Furthermore, we can simulate more intron gains than the empirical observation. Figure 9 shows the counts of exons at different percentage of intron gains via simulation (gaining introns from 0% to 124%). The simulation suggests that, at 100% intron gains, the number of exons satisfying exon recognition ([50 250]) peaks, so do exons falling in [75 200] for positioning nucleosome, which agrees with above observation on the minimum exon size (~70 nt) needed to efficiently position a mono-nucleosome. We also arbitrarily assume the maximum size of 200 nt given that the nucleosome occupancy over exon-intron boundary gets blurred when exons are longer than 200 nt (Figure 8). For the first time, the GRFP simulation suggests that the number of introns existing in *H. sapiens* genome provides the maximum percentage of exons for both satisfying the exon definition model and efficiently positioning a mono-nucleosome.

In summary, both exon definition and nucleosome positioning might play an important role in preventing further intron gains (or fragmentation of exons). Given that the concept of exon definition is proposed for vertebrate genomes, for genomes containing mostly short introns like *D. melanogaster*, chromatin structure might play a more important role in shaping the evolution of exon sizes. As a result, a large portion of its exons is longer than 400 nt and we have estimated that it has probably lost 60% of its introns over evolution [23].

## Conclusions

We have demonstrated that there is a strong correlation between exon size and exon GC content when flanking intron GC content is low. We find that both exon and flanking intron GC contents play a key role in determining how the exon-intron structure is marked by nucleosome. By simulation, we demonstrate that exon sizes are optimized over evolution for both exon recognition and nucleosome positioning. The correlation between nucleosome occupancy and exon size suggests that the chromatin organization plays an important role in shaping the evolution of exon-intron structure.

## Competing interests

The authors declare that they have no competing interests.

## Authors' contributions

LW designed and performed the data analyses. LW, LDS, and DW wrote the manuscript. All authors read and approved the final manuscript.

## Acknowledgements

We thank the National Science Foundation (DBI-0735191) and National Institute of Health (P41-HG02223) for funding aspects of this work.

# Figure

**Figure 1 Size distributions of exons in eight mutually exclusive categories.** The histograms of the exon sizes (in log scale) are fitted with a normal function (iuexon, itexon, ituexon, 5utexon, and 5uexon) or a mixture of two normal functions (each in dashed lines for iutexon, 3tuexon, and 3uexon). The horizontal line in each subplot indicates an intron, and the rectangle box indicates an exon. Coding fraction of the exon is shown in dark. The mean value and standard deviation are also shown for each fitted distribution.

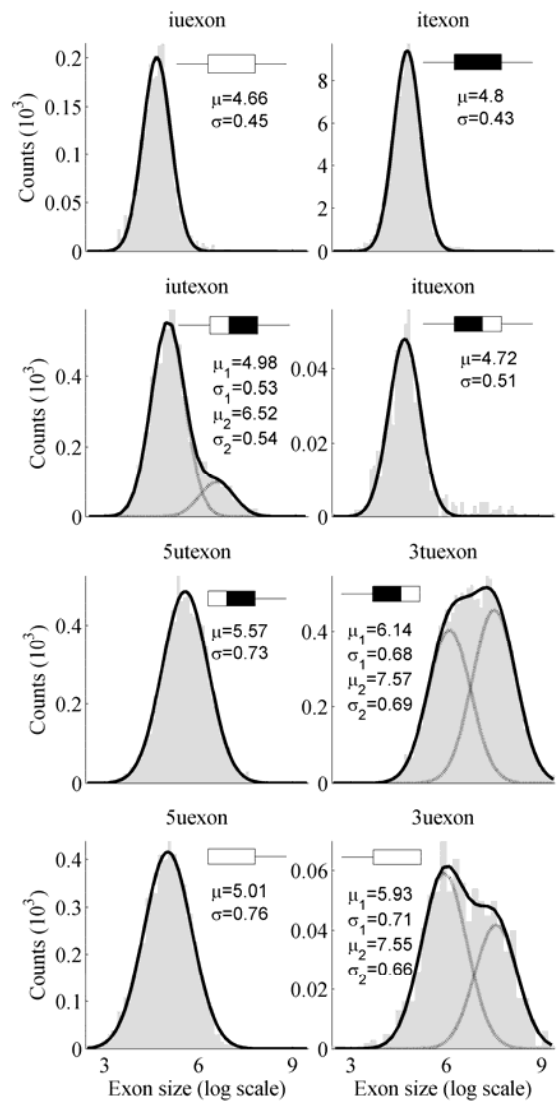

**Figure 2 Correlation of exon size with its GC content.** Solid (dashed) line shows the correlation without (with) considering GC content in flanking introns. Exons are grouped in GC windows centered from 0.3 to 0.7 with size of 0.05. For the dashed line, exons are also grouped by the GC content in flanking introns (shown in the legend). The exon size is calculated as the mean of the normal distribution fitted to the log scale distribution of exon size in each GC window (only the group with more than 500 exons are shown). The vertical error bar shows the standard error for each exon group.

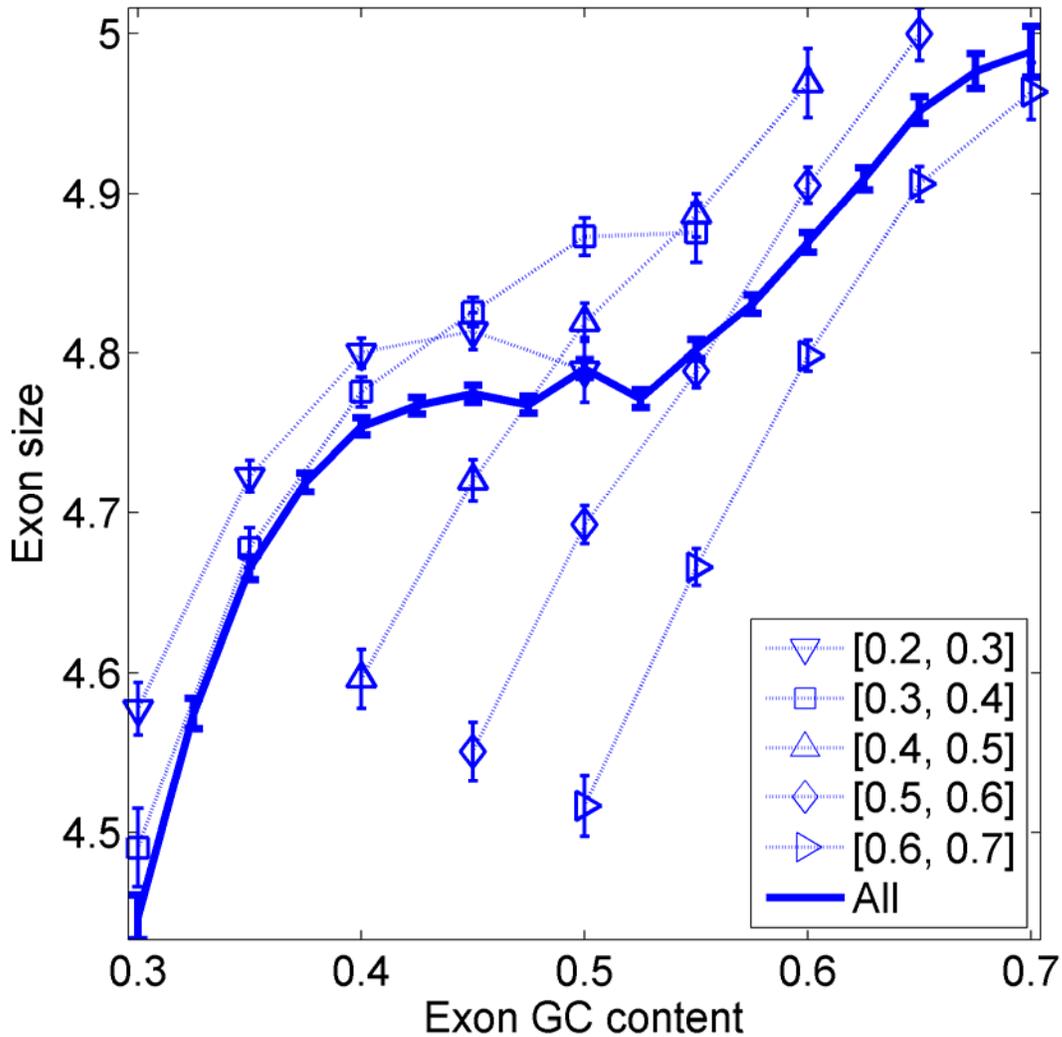

**Figure 3 Nucleosome occupancy based on direct sequencing of nucleosome ends in activated T cells.** Exons are aligned by their 3′ splice site (left) or their 5′ splice site (right). The splice site is marked at 0. All nucleosome occupancy values are subtracted by the average values between -120 nt and -70 nt and then averaged. The horizontal line below the curve indicates an intron, and the open rectangle box indicates an exon.

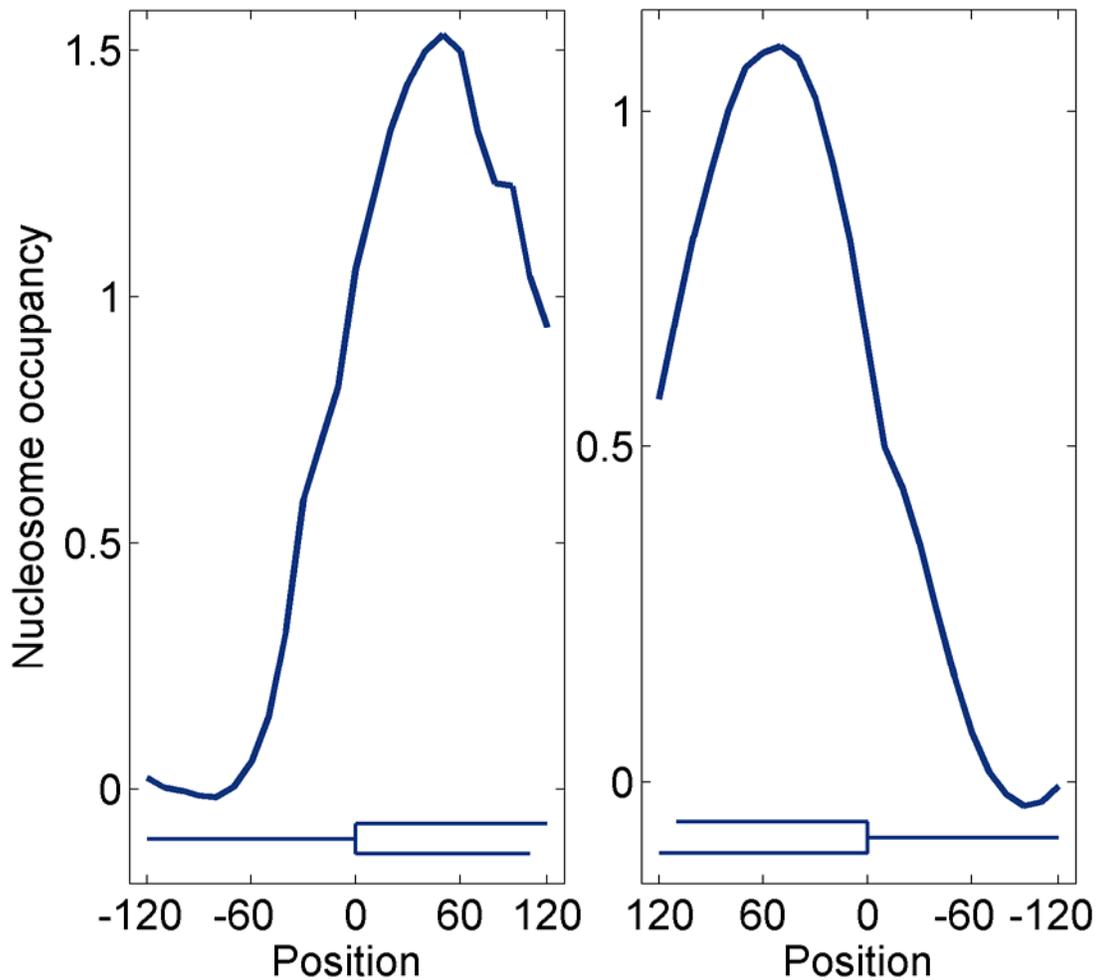

**Figure 4 Correlation of nucleosome occupancy with exon GC content.** Exons are grouped by the GC content in flanking introns (shown in the legend) and themselves with GC window size of 0.05 (only groups with more than 500 exons are shown). For each group, exons are aligned by their splice sites and the nucleosome occupancy values are calculated as the difference between exon and intron (see methods section for details).

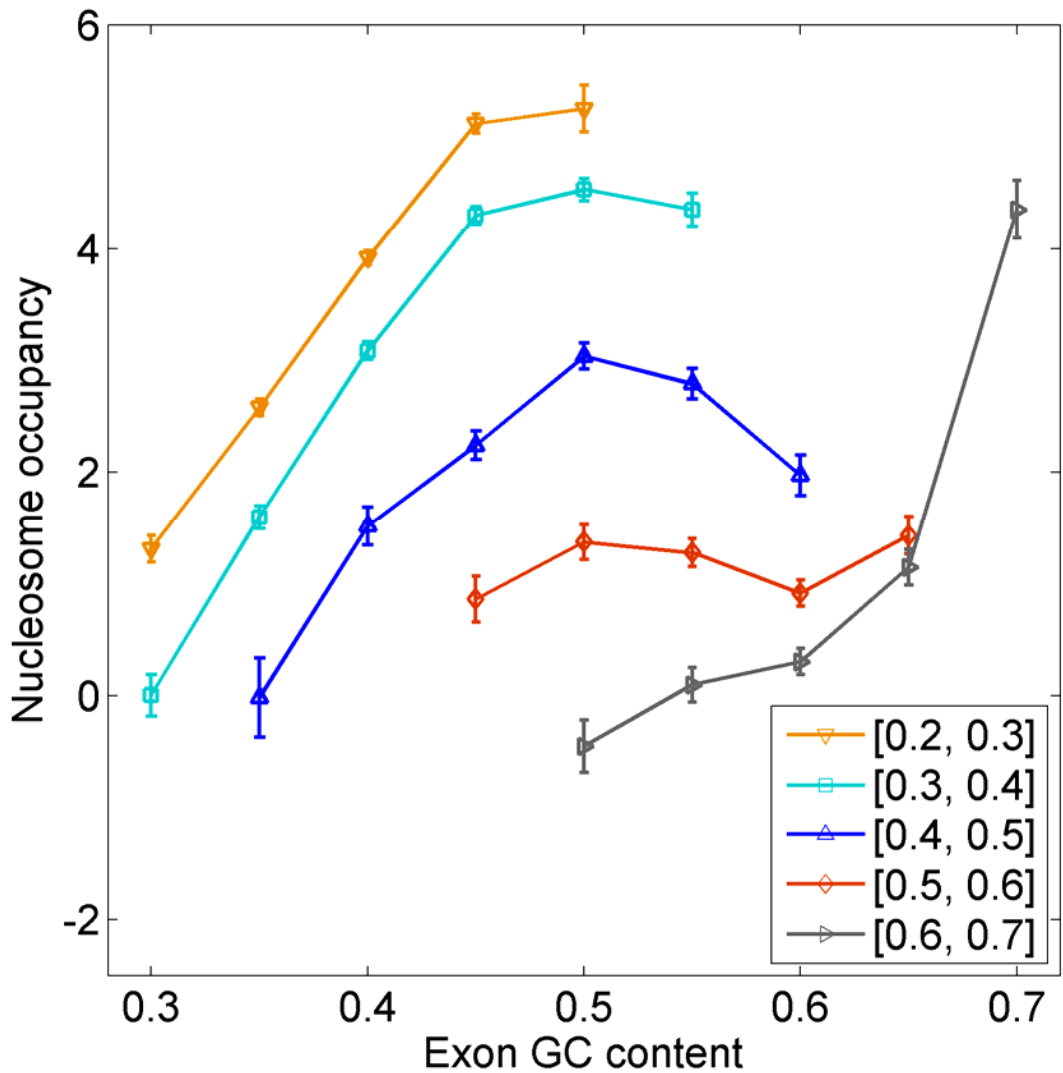

**Figure 5. Correlation of exon size with nucleosome occupancy.** Exons are grouped by the GC content in flanking introns (shown in the legend) and nucleosome occupancy windows centered from -10 to 14 with window size of 2. The average exon size is calculated from fitting a normal distribution to the log scale of exon size histogram in each group. The vertical error bar shows the standard error for each exon group. The horizontal dashed line shows where log(146) is, and 146 nt is the size of DNA fragment wrapping around a mono-nucleosome.

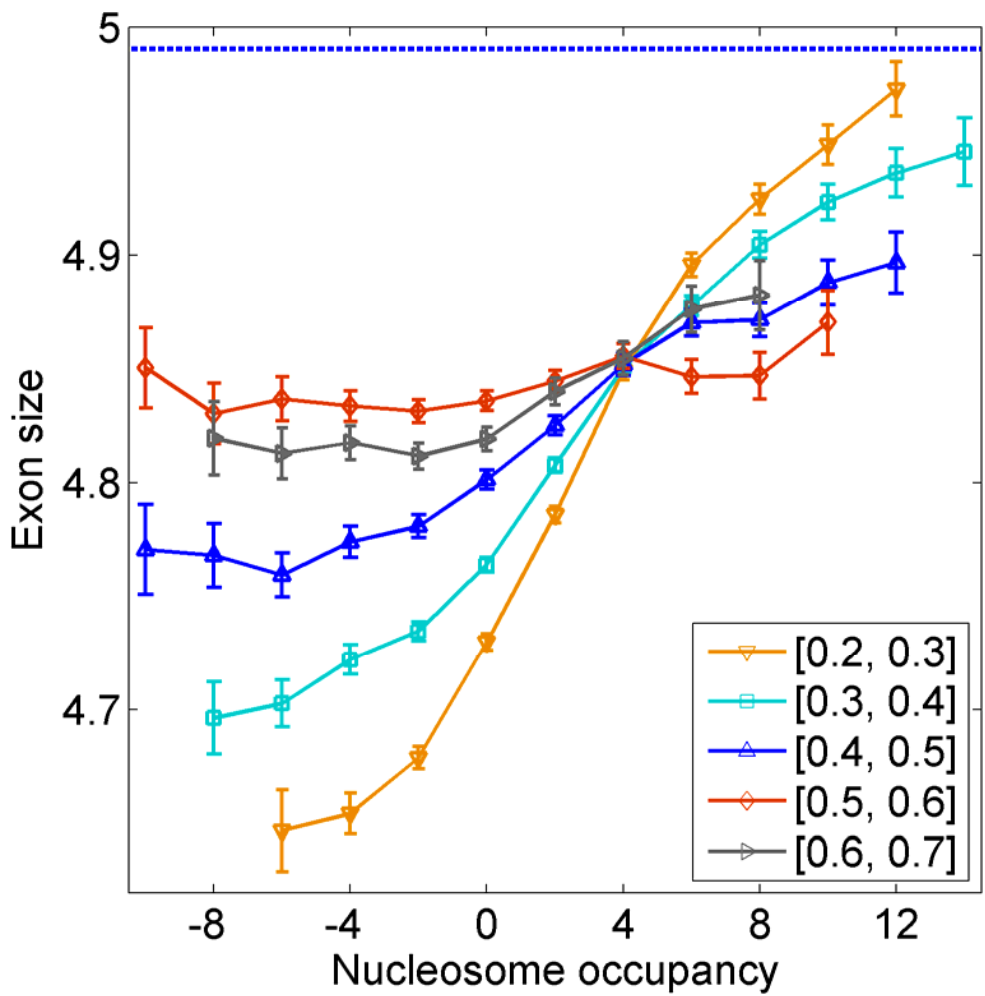

**Figure 6. Fitted exon size (mean) and standard deviation for exons with position relative to TSS (solid line) and TES (dashed line).** Exons are grouped by the distance to TSS or TES (<500 nt, 500-1000 nt, …, >10000 nt), which is calculated as the distance between the center of the exon and TSS (TES).

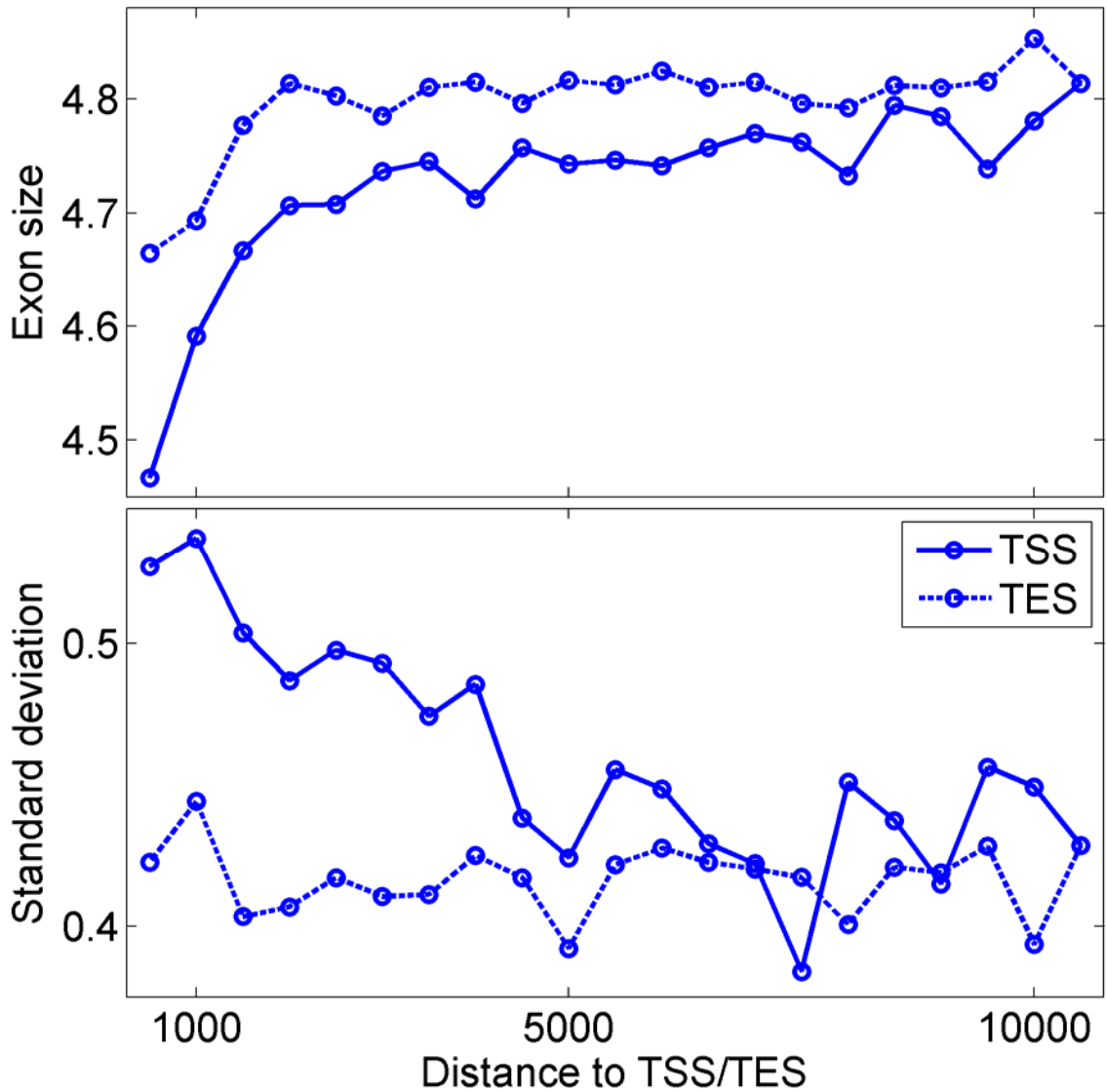

**Figure 7. Nucleosome occupancy for exons near TSS and TES.** Exons are grouped by the distance to TSS or TES, and aligned by their 3′ splice site (left) or their 5′ splice site (right). The splice site is marked at 0. The horizontal line below the curve indicates an intron, and the open rectangle box indicates an exon. The distance to TSS (TES) is calculated as the distance between the center of the exon and TSS (TES).

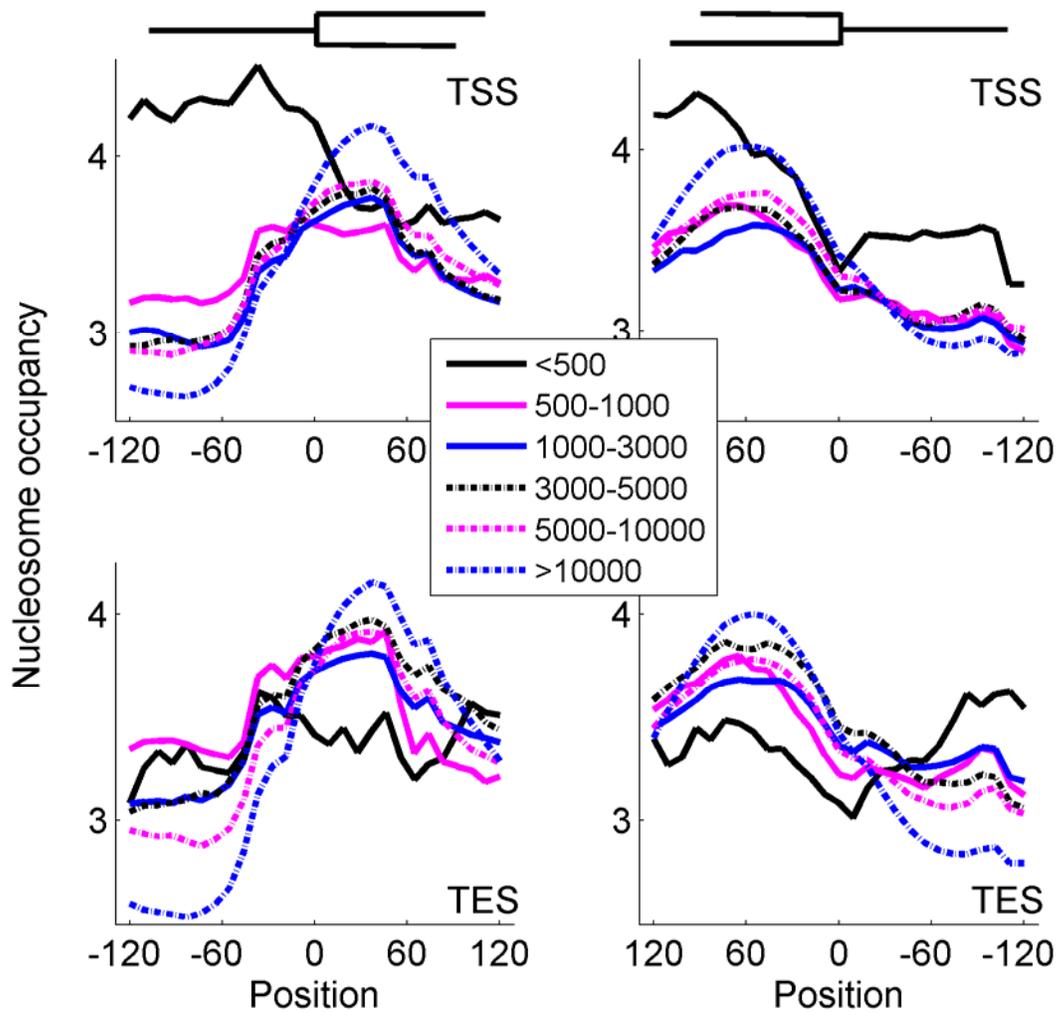

**Figure 8. Nucleosome occupancy density maps for exons and their flanking introns when grouped by exon size.** Exons with GC content higher than the GC content of flanking introns on either side are shown. Exons are grouped by size from 30 nt to 330 nt with a bin size of 10 (e.g. 30±5). The level of nucleosome occupancy is averaged in each group and then aligned by the 3′ splice site (position at 120).

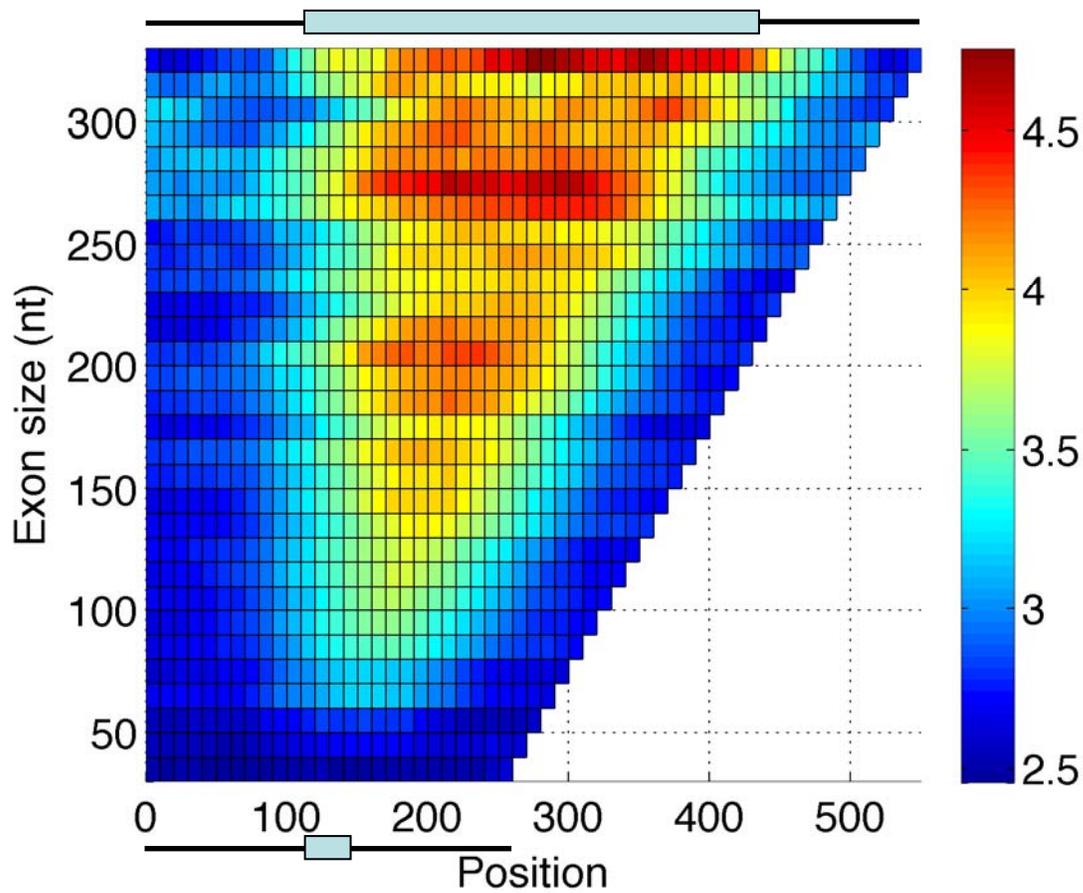

**Figure 9. Percentage of 'optimal' exons as a function of intron gains.** Intron gains are simulated with the GRFP model with the vertical line representing gaining 100% of the observed introns. The simulation (online at http://y2u.be/whja_hWvrOg) starts with one long exon with the length equal to the sum of all internal coding exon sizes (13256978 or e^16.4 for 104115 internal coding exons extracted from *H. sapiens*), exactly reproduces the empirical exon size distribution when 104114 introns (100%) get inserted, and stops after gaining extra 25000 introns (total 124%). The percentages of exons in each size range are counted after every 1000 intron gains, plotted, (the inset picture shows the overall trends), and labeled for the top 6 curves (zooming in). Curves under [75 200] are for percentages of exons with size falling in [80 200], [85 200], [90 200], [100 200], [110 200], [120 200], [130 200], [140 200], [150 200], and [160 200]. Curves labeled with [50 250] and [75 200] peak at 100% intron gains, suggesting that exons are optimized in sizes falling in these two ranges.

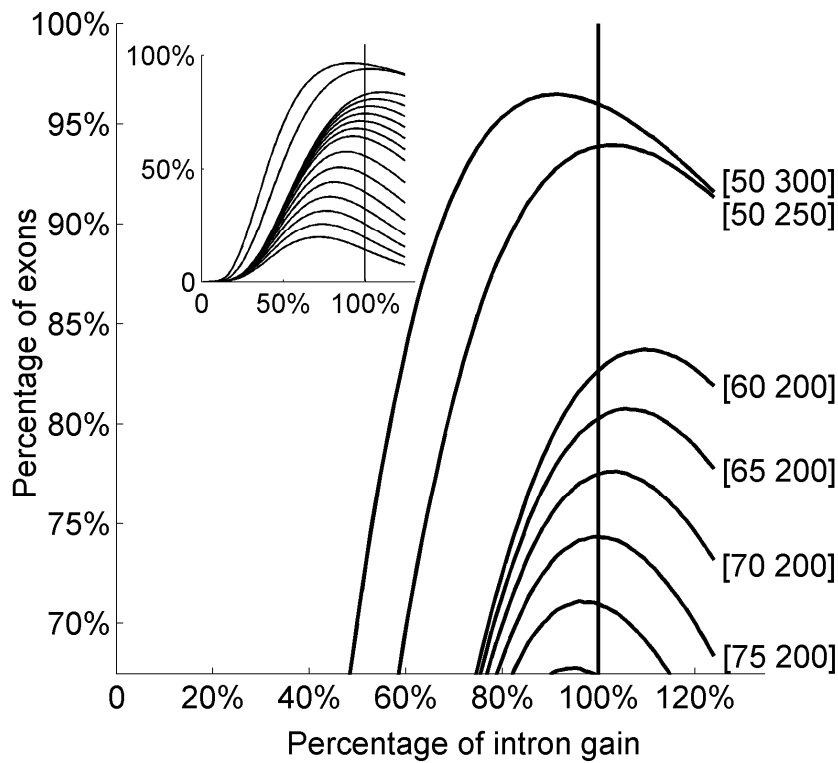